\documentclass[aps,prl,twocolumn,superscriptaddress]{revtex4}
\usepackage{graphicx}
\usepackage{dcolumn}
\usepackage{bm}
\usepackage{siunitx}
\usepackage{color}

\usepackage{graphicx}
\usepackage{grffile}

\usepackage{amsmath}
\usepackage{amssymb}
\usepackage{color}

\usepackage{booktabs}
\usepackage{tabularx}
\usepackage{subfig}
\usepackage{caption}

\usepackage[T1]{fontenc}
\usepackage{bm,bbold}

\usepackage[english]{babel}
\usepackage[utf8]{inputenc}
\usepackage{multirow}

\newcommand{\bra}[1]{\big<#1\big|}
\newcommand{\ket}[1]{\big|#1\big>}

\newcommand{\bk}{\mathbf{k}}
\newcommand{\bkp}{\mathbf{k'}}

\newcommand{\bq}{\mathbf{q}}
\newcommand{\bqp}{\mathbf{q'}}

\newcommand{\hb}{\bar{h}}
\newcommand{\eb}{\bar{e}}
\newcommand{\hbp}{\bar{h}'}
\newcommand{\ebp}{\bar{e}'}

\def\dz2{d$_{\text{z}^2}$}

\def\dx2y2{d$_{\text{x}^2\text{y}^2}$}
\def\G0W0{G$_0$W$_0$}
\def\scGW0{scGW$_0$}

\begin{document}

\bibliographystyle{naturemag}

\title{Biexciton fine structure in monolayer transition metal dichalcogenides}

\author{Alexander Steinhoff}
\email{asteinhoff@itp.uni-bremen.de}

\affiliation{Institut f\"ur Theoretische Physik, Universit\"at Bremen, P.O. Box 330 440, 28334 Bremen, Germany}

\author{Matthias Florian}

\affiliation{Institut f\"ur Theoretische Physik, Universit\"at Bremen, P.O. Box 330 440, 28334 Bremen, Germany}

\author{Akshay Singh}

\affiliation{University of Texas at Austin, Texas}
\affiliation{Present address: Department of Material Science and Engineering, Massachusetts Institute of Technology, Cambridge, MA 02139, USA}

\author{Kha Tran} 

\affiliation{University of Texas at Austin, Texas}

\author{Mirco Kolarczik}

\affiliation{Institut f\"ur Optik und Atomare Physik, Technische Universit\"at Berlin, Berlin, Germany}

\author{Sophia Helmrich}

\affiliation{Institut f\"ur Optik und Atomare Physik, Technische Universit\"at Berlin, Berlin, Germany}

\author{Alexander W. Achtstein}

\affiliation{Institut f\"ur Optik und Atomare Physik, Technische Universit\"at Berlin, Berlin, Germany}

%

%

\author{Ulrike~Woggon}

\affiliation{Institut f\"ur Optik und Atomare Physik, Technische Universit\"at Berlin, Berlin, Germany}

\author{Nina Owschimikow}

\affiliation{Institut f\"ur Optik und Atomare Physik, Technische Universit\"at Berlin, Berlin, Germany}

\author{Frank Jahnke}

\affiliation{Institut f\"ur Theoretische Physik, Universit\"at Bremen, P.O. Box 330 440, 28334 Bremen, Germany}
\affiliation{MAPEX Center for Materials and Processes, Universit\"at Bremen, 28359 Bremen, Germany}

\author{Xiaoqin Li}

\affiliation{University of Texas at Austin, Texas}


\begin{abstract}


The optical properties of atomically thin transition metal dichalcogenide (TMDC) semiconductors are shaped by 
the emergence of correlated many-body complexes due to strong Coulomb interaction. 
Exceptional electron-hole exchange predestines TMDCs to study fundamental and applied properties of Coulomb complexes such as 
valley depolarization of excitons and fine-structure splitting of trions.
Biexcitons in these materials are less understood and it has been established only recently that they are spectrally located between exciton and trion. 

Here we show that biexcitons in monolayer TMDCs exhibit a distinct fine structure on the order of meV due to electron-hole exchange.
Ultrafast pump-probe experiments on monolayer WSe$_2$ reveal 
decisive biexciton signatures and a fine structure in excellent agreement with a microscopic theory.
We provide a pathway to access biexciton spectra with unprecedented accuracy, which is valuable beyond the class of TMDCs, 
and to understand even higher Coulomb complexes under the influence of electron-hole exchange.

\end{abstract}

\maketitle


The exchange interaction between electrons and holes in semiconductors 
has long been a major field of interest in both theory and experiment.
It has been discussed 
in the context of longitudinal-transverse exciton splitting in bulk materials \cite{onodera_excitons_1967, denisov_longitudinal_1973} and of exciton fine-structure splitting in quantum dots, where it is a major limitation to the generation of entangled photons \cite{warming_hole-hole_2009,kadantsev_theory_2010}. 
Biexcitons, also known as exciton molecules, are fundamental to the nonlinear optical response of semiconductors. Based on a many-body theory including biexcitonic effects, nonlinear pump-probe experiments in InGaAs quantum wells have been successfully described~\cite{sieh_coulomb_1999, schafer_semiconductor_2002} 
without taking electron-hole exchange into account. On the other hand, several authors have recognized the relevance of electron-hole exchange for biexcitons in materials like CuCl and CdS
\cite{forney_electron-hole_1974, quattropani_biexcitons_1975, ekardt_influence_1976, bassani_biexciton_1976, quattropani_theory_1977, chung_theory_1983, ungier_electron-hole_1989}.
Although in some cases significant corrections of biexciton binding energies due to electron-hole exchange have been reported, a fine-structure splitting of biexcitons has not been discussed. 
\par Atomically thin TMDC semiconductors 
offer new possibilities for studying Coulomb correlation effects by introducing reciprocal-space valleys as a new degree of freedom selectively addressable by circularly polarized light.
Electron-hole exchange has strong implications for the valley dynamics and is currently an object of intense research \cite{yu_valley_2014, yu_dirac_2014, glazov_spin_2015, schmidt_ultrafast_2016}. 
The theory of electron-hole exchange in TMDC semiconductors has been put on a microscopic basis by Qiu et al. \cite{qiu_nonanalyticity_2015}, who discussed the resulting non-analytic exciton dispersion and strong splitting between 
bright and dark excitons. Similarly strong effects are observed in the fine-structure splitting of inter- and intra-valley trions in these materials \cite{jones_excitonic_2016, plechinger_trion_2016, singh_long-lived_2016}. 
It is worth speculating whether biexcitons could also have a fine structure with implications for nonlinear optical effects utilized in coherent control schemes~\cite{voss_coherent_2006} and for manipulation of exciton spin coherences~\cite{gilliot_measurement_2007}.
However, even modern variational approaches that have been applied to describe biexcitons in TMDC semiconductors rely on simple band structures and Coulomb interaction models, neglecting electron-hole exchange effects entirely \cite{mayers_binding_2015, zhang_excited_2015, kylanpaa_binding_2015, szyniszewski_binding_2017}.

\par In this paper, we provide evidence for a pronounced
biexciton fine structure in TMDC semiconductors by performing for the first time material-realistic calculations of biexciton spectra including electron-hole exchange interaction. 
We condense the full microscopic theory into a simple four-particle-configuration model for biexcitons to identify the relevant interaction processes and optical selection rules.
Our theory shows that the biexciton fine structure is caused by nonlocal electron-hole exchange, while local exchange leads to an increase of the binding energy of the lowest biexciton state.
Ultrafast pump-probe experiments performed on monolayer WSe$_2$ verify the theoretical prediction by showing a distinct biexciton fine structure in excellent quantitative agreement with the calculated spectrum.
All biexciton resonances are found to be spectrally located between exciton and trion, which is consistent with the spectral ordering observed in monolayer MoSe$_2$ by two-dimensional coherent spectroscopy \cite{hao_neutral_2017}.
Our results give rise to a new picture of multi-exciton states as quantum superpositions of many-body configurations dominating nonlinear optical response in TMDC semiconductors. Thereby we aim to stimulate new experiments to observe fine structures also in higher Coulomb complexes.

\begin{figure*}[h!t]
\begin{center}
\includegraphics[width=.8\textwidth]{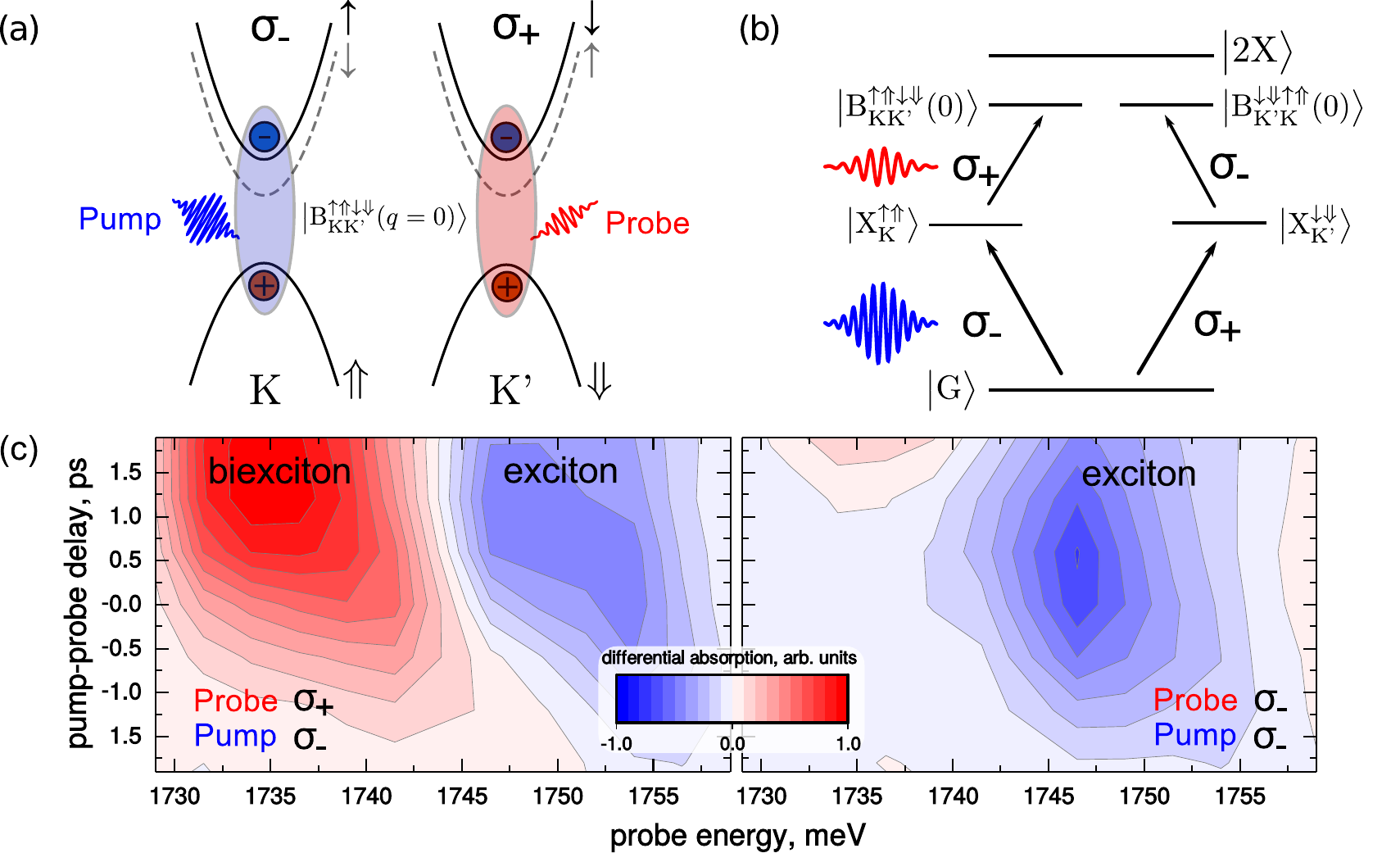}
\caption{\textbf{Pump-probe selection rules of biexcitons in monolayer TMDC.} \textbf{(a)} Coherent biexcitons with zero total momentum are resonantly excited by pump and probe photons with cross-circular polarization. The solid parabola denote the relevant like-spin conduction and valence bands around the K and K' points, while the dashed parabola denote split-off conduction bands with unlike spins. 
The biexciton state $\ket{\textrm{B}^{\uparrow\Uparrow\downarrow\Downarrow}_{\textrm{KK'}}(q=0)}$ emerges from the mutual Coulomb interaction of two electrons and two holes and can be pictured as being composed of two excitons in the K and K' valleys with momentum $\bq=0$. 
\textbf{(b)} Schematic of a pump-probe experiment with two equivalent excitation pathways leading to two equivalent bright biexciton states. The biexciton binding energy is given by the difference between the biexciton energy and the exciton-exciton scattering continuum edge denoted by $\ket{2\textrm{X}}$. 
\textbf{(c)} Differential absorption of WSe$_2$ pumped at the red wing of its neutral exciton resonance in a pump-probe experiment with cross- (left panel) and co-circularly polarized (right panel) pump and probe pulses.}
\label{fig:selection_rules}
\end{center}
\end{figure*}
%
\par Biexcitons can be detected in nonlinear optical measurements, where biexciton formation can either take place coherently or incoherently. In the latter case, the semiconductor is excited above the single-particle band gap to generate hot carriers that may relax and form excitons and biexcitons by passing excess energy to the lattice before carrier recombination sets in. 
In a resonant pump-probe experiment, a pump tuned to the exciton energy leads to an \textit{induced} absorption of the probe, tuned to the energy red-shifted from the pump by the biexciton binding energy.\cite{schafer_semiconductor_2002} This coherent process of biexciton creation is cleaner and easier to compare with a microscopic theory. 
The measured pump-probe signal is governed by optical selection rules that typically allow the 
observation of biexcitons only for certain combinations of pump- and probe-polarizations. 
In atomically thin TMDC semiconductors, K and K' valleys are linked by time-reversal symmetry. Further, strong spin-orbit splitting of the valence band leads to a coupling of valley and spin selection rules, 
and thus valleys can be selectively excited with circularly polarized light. \cite{xu_spin_2014}
As it is known from conventional semiconductors, biexcitons can only be excited when pump and probe photons have opposite polarization
\cite{sie_intervalley_2015} as illustrated in Fig.~\ref{fig:selection_rules}. For co-circular pump-probe, no biexciton signal is expected.
To test this prediction, we perform near-resonant pump-probe experiments on a mechanically exfoliated tungsten diselenide (WSe$2$) monlayer kept at a temperature of 13~K in a closed-cycle cryostat. For details of the experiment see the Methods section.
The results of measurements with cross- and co-circular polarization of pump and probe pulses are shown in the left and right panels of Fig.~\ref{fig:selection_rules}(c), respectively. As will be shown later, biexcitons in this material show up in the spectral region between the neutral exciton and the trion. Indeed, the experiment shows a distinct induced absorption in this area, which is only present for cross-circular polarization of pump and probe pulses. Spectra of the sample and a set of pump-probe experiments over an extended spectral 
and time range are shown in the Supplementary Information.

\par
\begin{figure*}[h!t]
\begin{center}
\includegraphics[width=\textwidth]{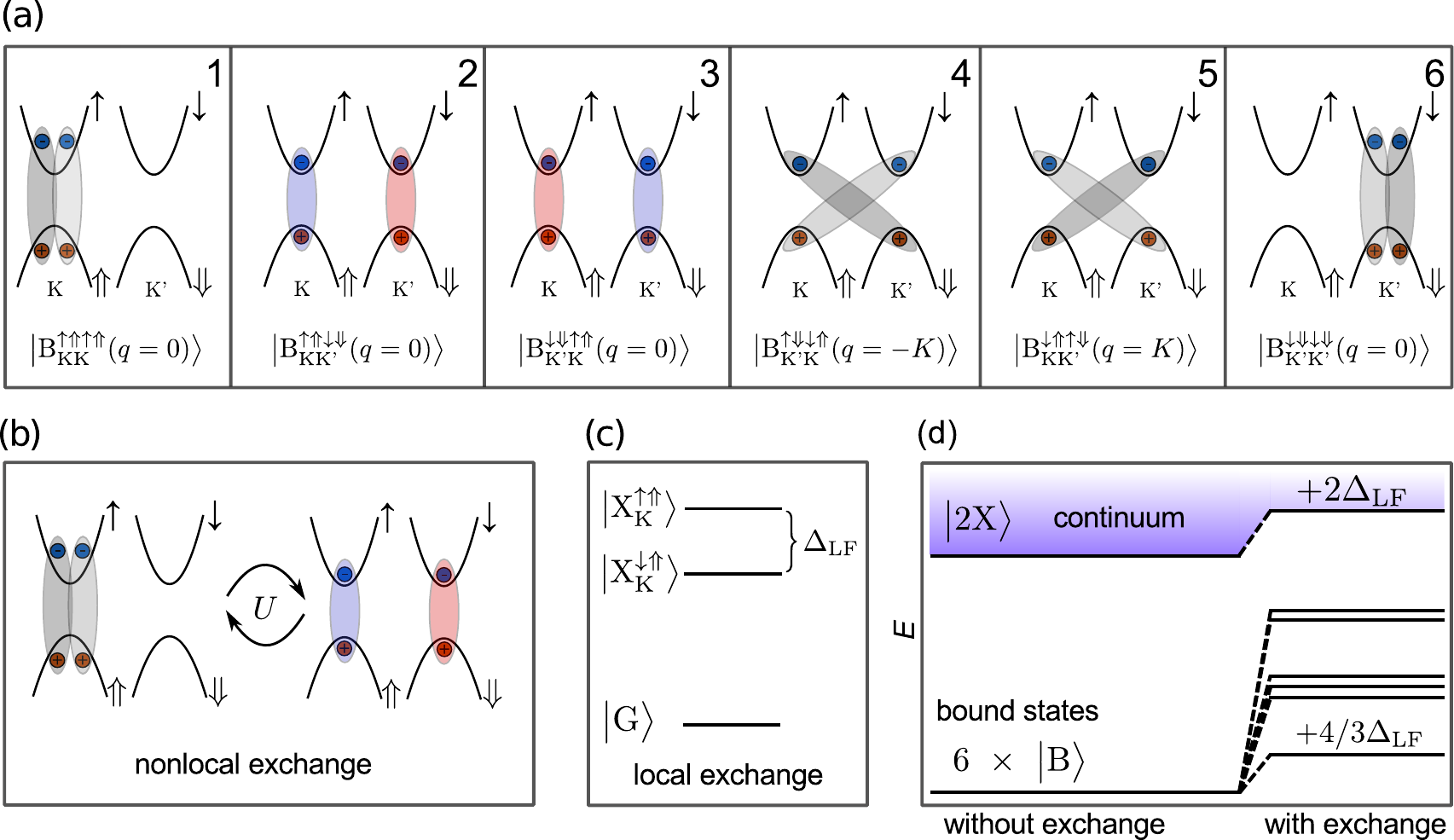}
\caption{\textbf{Configuration picture of biexcitons in monolayer WSe$_2$.} 
\textbf{(a)} Biexciton configurations realized by pairing like- and unlike-spin excitons with different total momenta $\bq$ in monolayer TMDC. Only 
two of the configurations (2 and 3) can be optically excited in a pump-probe experiment as shown in Fig.~\ref{fig:selection_rules}.
\textbf{(b)} Nonlocal electron-hole exchange induces interaction, given by the matrix element $U$, between biexciton configurations that differ by the spin flip of one electron-hole pair. 
\textbf{(c)} Local electron-hole exchange induces a blue shift $\Delta_{\textrm{LF}}$ of like-spin (triplet) excitons leading to a singlet-triplet exciton splitting. The blue shift is transferred to biexciton configurations composed of two like-spin excitons (configurations 1,2,3 and 6). 
\textbf{(d)} Schematic four-particle energy spectrum including six degenerate biexciton configurations as introduced in (a) below the exciton-exciton scattering continuum. Local electron-hole exchange interaction shifts the continuum by $2\Delta_{\textrm{LF}}$, while nonlocal exchange leads to a fine structure of biexciton states by mixing all configurations. The shift of the lowest biexciton state can be qualititively understood from the configuration-interaction model, while precise numbers are obtained from a microsopic theory, see discussion in the text.}
\label{fig:eh_exchange}
\end{center}
\end{figure*}

For a quantitative description of the observed absorption feature, we introduce
a simple yet powerful model based on a small Hilbert space of four-particle configurations to analyze the effect of electron-hole exchange on optical biexciton spectra. The model is underpinned by a microscopic theory of biexcitons, which we discuss in detail in the Methods section and use to obtain precise numbers that can be compared to experiment. In pump-probe experiments, only biexcitons with vanishing total momentum can be observed. Due to translational symmetry of the crystal, biexcitons with different total momentum do not couple, so that we can limit the Hilbert space of our model to biexcitons with vanishing total momentum. Moreover, monolayer WSe$_2$ exhibits a conduction-band splitting of about $50$ meV at the K-point due to spin-orbit interaction as schematically shown in Fig.~\ref{fig:selection_rules}(a).
Due to the large conduction-band splitting, coupling between upper and lower conduction band is expected to be weak.
Hence we can exclude the lower conduction band from our theory, which significantly simplifies the discussion of results.
Consequently, in the subspace of zero-momentum biexcitons there are six degenerate configurations 
including biexcitons composed of two dark excitons with opposite finite momentum $|\textbf K- \textbf K'|$ presented in Fig.~\ref{fig:eh_exchange}(a):
$\Big\{\ket{\textrm{B}^{\uparrow\Uparrow\uparrow\Uparrow}_{\textrm{KK}}(0)},\ket{\textrm{B}^{\uparrow\Uparrow\downarrow\Downarrow}_{\textrm{KK'}}(0)},\ket{\textrm{B}^{\downarrow\Downarrow\uparrow\Uparrow}_{\textrm{K'K}}(0)}  ,$ $\ket{\textrm{B}^{\uparrow\Downarrow\downarrow\Uparrow}_{\textrm{K'K}}(-K)}, \ket{\textrm{B}^{\downarrow\Uparrow\uparrow\Downarrow}_{\textrm{KK'}}(K)},\ket{\textrm{B}^{\downarrow\Downarrow\downarrow\Downarrow}_{\textrm{K'K'}}(0)} \Big\}$.
%
Only the configurations $\ket{\textrm{B}^{\uparrow\Uparrow\downarrow\Downarrow}_{\textrm{KK'}}(0)}$ and $\ket{\textrm{B}^{\downarrow\Downarrow\uparrow\Uparrow}_{\textrm{K'K}}(0)}$ corresponding to the pump-probe excitation pathways shown in Fig.~\ref{fig:selection_rules}(b) have non-zero oscillator strength
Nevertheless, all six configurations are eigenstates with energy $E$ of the four-particle Hamiltonian $H_0$ that contains the kinetic energies of two electrons and two holes as well as the direct Coulomb interaction between them in analogy to the hydrogen molecule. 
By adding electron-hole exchange to this picture, we introduce interaction between the configurations leading to new eigenstates and -energies. This concept of interacting configurations is well-known from many-electron atoms or semiconductor quantum dots. It is the new eigenstates including interaction which are observed in experiment instead of the configurations. 
All six configurations are coupled by electron-hole exchange and are thus necessary to be included in the model.
Microscopically, there are local and nonlocal contributions to electron-hole exchange. Nonlocal exchange mixes excitons in the K and K' valleys, thereby reducing valley 
selectivity \cite{yu_valley_2014, yu_dirac_2014, glazov_spin_2015, schmidt_ultrafast_2016}. At the same time, a non-analytic lightlike exciton dispersion for like-spin excitons emerges together with a momentum-dependent splitting. Local exchange on the other hand leads to an overall blue shift of like-spin excitons, which we refer to as local-field effect. A detailed discussion for monolayer MoS$_2$ is given in Ref.~\cite{qiu_nonanalyticity_2015} and comparable numerical results for monolayer WSe$_2$ are shown in Fig.2 in the Supporting Information.
The effects of electron-hole exchange on biexciton configurations are schematically explained in Fig.~\ref{fig:eh_exchange}(b) and (c). The nonlocal and local contributions to exchange are described by the Hamiltonians $H_\textrm{U}$ and $H_{\textrm{LF}}$, respectively, which have to be added to $H_0$. We set up the total Hamiltonian by using the six configurations as a basis:
\begin{equation}
    \begin{split} 
      H_{ij}&=\bra{B_i}H_0\ket{B_j}+\bra{B_i}H_\textrm{U}+H_{\textrm{LF}}\ket{B_j} \\ &=E+U_{ij}+\Delta_{\textrm{LF},ij}\,.
        \label{eq:hamilton}
    \end{split} 
\end{equation}
%
%
\begin{figure*}
\begin{center}
\includegraphics[width=0.8\textwidth]{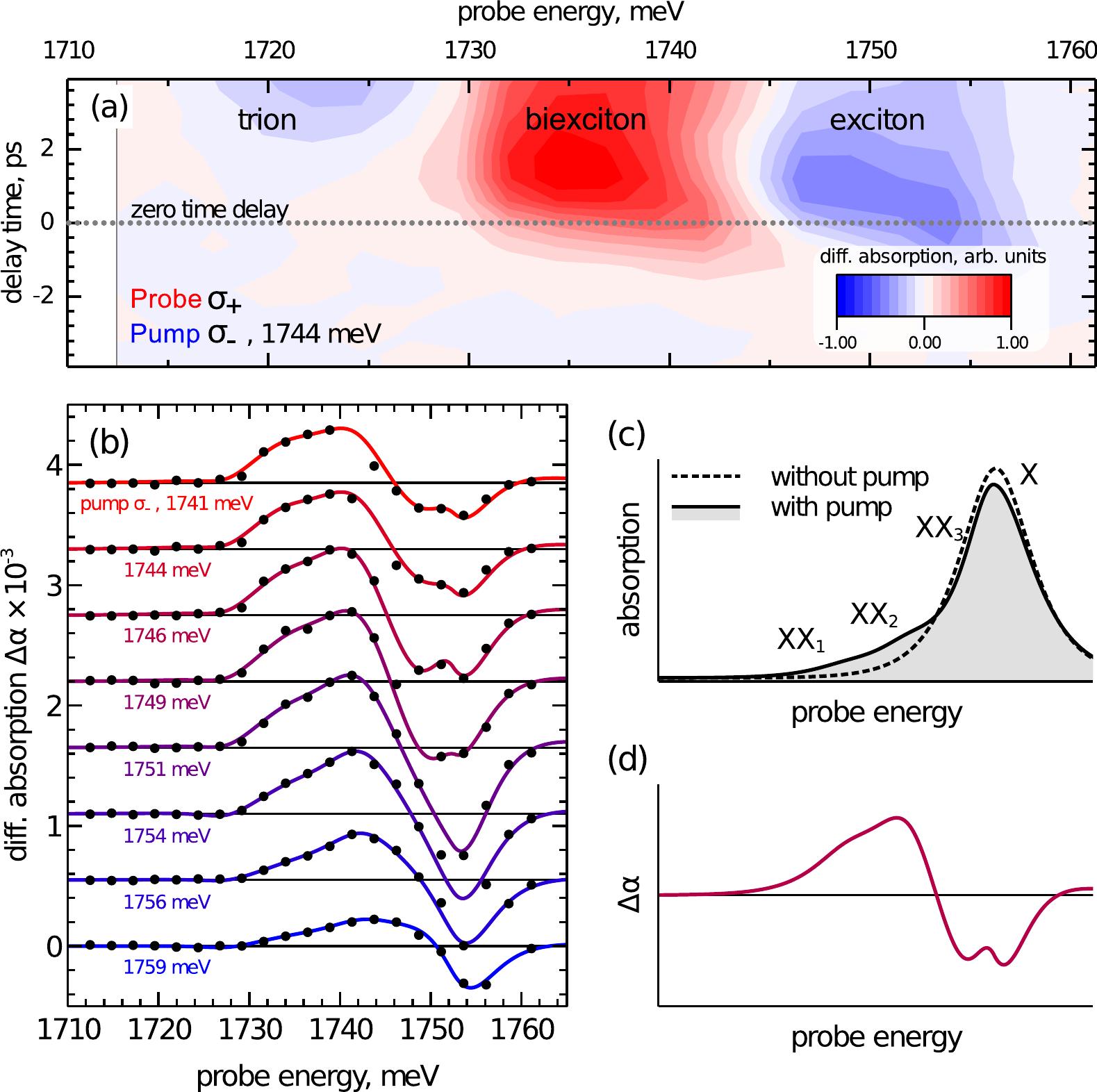}
\caption{\textbf{Differential absorption of monolayer WSe$_2$ on sapphire from theory and experiment.} \textbf{(a)} Time-resolved differential absorption data from cross-polarized pump-probe spectroscopy with a pump energy of 1744 meV. 
\textbf{(b)} 
The corresponding differential absorption spectra at zero time delay (filled black circles)
for pump-pulse energies ranging from 1759~meV (bottom) to 1741~meV (top) are compared to fits based on the predictions of our theoretical model (solid lines). Details on the fits are given in the text. The different data sets are vertically shifted for better visibility, with the zero differential absorption for each pump spectra indicated as thin lines. \textbf{(c)} Fit function for absorption spectra without and with pump at 1746~meV. The pump-induced emergence of asymmetry of the exciton line (X) as well as two prominent biexciton resonances (XX$_1$, XX$_2$) are visible. The third biexciton (XX$_3$) has a much smaller oscillator strength. \textbf{(d)} Differential absorption from the fitting curves in (c). Asymmetric lineshape, bleaching and shift of the exciton lead to the remarkable signature in $\Delta\alpha$ that is also observed in experiment.}
\label{fig:exp_theo}
\end{center}
\end{figure*}
The off-diagonal matrix elements take into account the spin-selection rules due to electron and hole spins in the configurations.
To obtain an analytically solvable problem, we assume that all non-zero exchange matrix elements have the same values $U$ and $\Delta_{\textrm{LF}}$, respectively. The full Hamiltonian is given in the Methods section.
To be more transparent, we first analyze the biexciton spectrum in the absence of local-field effects by diagonalizing $H_0+H_\textrm{U}$, obtaining
the eigenvalues 
\begin{equation}
    \begin{split} 
\Big\{E_1&=E,\,E_2=E+2U,\,E_3=E+2U, \\ E_4&=E+2U,\,E_5=E+4U,\,E_6=E+6U \Big\}\,. 
        \label{eq:eigenvalues}
    \end{split} 
\end{equation}
Hence without local-field effects the lowest 
biexciton energy, $E$, remains unchanged by exchange interaction. At the same time a triplet of states is split off by twice the exchange matrix element $U$.
Including local-field effects perturbatively via $H_{\textrm{LF}}$, we find the first-order energy correction to the lowest eigenenergy to be $4/3 \Delta_{\textrm{LF}}$.
The lowest eigenstate is thus shifted less than the exciton-exciton scattering continuum ($2 \Delta_{\textrm{LF}}$) as illustrated in Fig.~\ref{fig:eh_exchange}(d).
This is due to the fact that the unlike-spin biexciton configurations 4 and 5 in Fig.~\ref{fig:eh_exchange}(a) are mixed into the new eigenstates but do not experience any local-field shift. 
The net result is an increase of the largest binding energy by $2/3 \Delta_{\textrm{LF}}$ in good agreement with the full microscopic solution.
\par
We draw conclusions about the optical visibility of the interacting states under cross-circular excitation by calculating dipole matrix elements:
\begin{equation}
    \begin{split} 
      |\textbf d_i|^2&=\left|\bra{\phi_i}\sigma_{+}\ket{\textrm{X}^{\uparrow\Uparrow}_{K}}\right|^2 \\ &=\left|\bra{\phi_i}\textrm{B}^{\uparrow\Uparrow\downarrow\Downarrow}_{\textrm{KK'}}(0)\big>+\bra{\phi_i}\textrm{B}^{\downarrow\Downarrow\uparrow\Uparrow}_{\textrm{K'K}}(0)\big>    \right|^2 \,,
        \label{eq:dipole}
    \end{split} 
\end{equation}
where $\ket{\phi_i}$ are the eigenstates 
including nonlocal electron-hole exchange
and $\sigma_{+}$ creates an electron-hole pair at K' as shown in Fig~\ref{fig:selection_rules}~(b). It follows that three of the biexciton states \textit{including} exchange interaction are bright with $|\textbf d _1|^2=2/3$, $|\textbf d _2|^2=1$ and $|\textbf d _6|^2=1/3$ corresponding to the eigenenergies $E_1$, $E_2$ and $E_6$ given above.
This remains true in the presence of local-field effects although the values of dipole matrix elements change. Hence we expect to observe three bound biexcitons, which are quantum superpositions of the basic four-particle configurations, with different oscillator strengths.






\par
The above prediction of a biexciton fine structure is confirmed by a quantitative analysis of the experimental data.
In Fig.~\ref{fig:exp_theo}(a) we present time-resolved differential absorption data obtained from the cross-polarized pump-probe signal exhibiting distinct signatures of exciton and trion as well as an additional resonance between them. In Fig.~\ref{fig:exp_theo}(b), the corresponding differential absorption spectra at zero time delay are shown. The
pump energy is varied across the exciton resonance from 1759~meV to 1741~meV while the probe wavelength is scanned from 1762~meV down to 1712~meV.
Biexciton signatures are expected to show up as positive signals, as they result from induced absorption. 
The experimental data can be perfectly described by fitting curves that take into account three biexciton resonances with binding energies 18.3$\pm$0.5 meV, 10.3$\pm$0.5 meV and 3 meV, respectively. For the fitting, while the amplitudes of each biexciton resonance are allowed to vary with changing pump-pulse energies, the center biexciton consistently has the strongest amplitude as estimated from our configuration model. 
For further details of the fitting process see the Supporting Information.
In addition to the biexcitons, our fit takes into account the differential exciton absorption resonance. We allow for a pump-induced energy shift, absorption bleaching and asymmetry of the line shape, where the lines becomes narrower on the red side and wider on the blue side. The latter is required to understand the 
signatures between 1750 and 1755 meV that can be fit systematically for all pump wavelengths, which we illustrate for an example in Fig.~\ref{fig:exp_theo}(b) and (c). The exciton line asymmetry can be attributed to dephasing processes due to biexciton scattering states as discussed in \cite{hulin_excitonic_1990, schafer_semiconductor_2002}. Finally, we account for a differential trion signal at 25 meV below the exciton by an additional negative resonance. As fit model we use Sech-functions for all resonances \cite{esser_photoluminescence_2000} with a width (FWHM) of the exciton line without pump of 10.5 meV and the same for all biexciton lines. In most cases, we extract a small exciton redshift below 1 meV, which points to dominant biexcitonic effects on the line shift \cite{combescot_excitonic_1988,schafer_semiconductor_2002}. 

\par
After comparing the predictions of our configuration model with differential absorption spectra on a qualitative level, we further substantiate our results by a quantitative assessment of biexciton binding energies.
A full microscopic description of coherent biexcitons
is obtained from a generalized four-particle Schr\"odinger equation in reciprocal space including realistic band structures and Coulomb matrix elements. Electron-hole exchange interaction can be systematically taken into account, which is beyond modern variational approaches applied to TMDC semiconductors. \cite{mayers_binding_2015, zhang_excited_2015, kylanpaa_binding_2015, szyniszewski_binding_2017} 
By diagonalizing the 
full microscopic Hamiltonian
we obtain energies of bound and scattering four-particle states. Biexciton binding energies ($E_{\textrm{B}} = E_{\textrm{XX}} - 2E_{\textrm{X}}$) are then extracted by comparing the lowest eigenvalues of the full four-body problem ($E_{\textrm{XX}}$) to the exciton-exciton continuum edge ($2E_{\textrm{X}}$) from a calculation for two independent excitons.
Neglecting electron-hole exchange, the calculation yields a binding energy of 19.5 meV for the biexciton that is shown in Fig.~\ref{fig:selection_rules}~(a) in freestanding monolayer WSe$_2$. 
We compare this result to previous variational calculations in the Supporting Information.
Including electron-hole exchange, we obtain six bound biexciton states with binding energies $\big\{24.2,16.4,15.6,15.6,7.1,5.5\big\}$ meV. Hence the largest binding energy 
increases
by about 4.5 meV due to exchange. This underlines the quality of our configuration model, which predicts an increase of biexciton binding energy due to local-field effects by $2/3 \Delta_{\textrm{LF}}$, where we find $\Delta_{\textrm{LF}} = 7$ meV from a microscopic calculation.
The correction of biexciton binding energies being proportional to the singlet-triplet exction splitting $\Delta_{\textrm{LF}}$ has been discussed before for materials like CuCl and CdS \cite{quattropani_theory_1977}.
At the same time, a group of three almost degenerate less tightly bound states as well as two isolated states emerge due to nonlocal electron-hole exchange, again in agreement with our configuration model. This spectral structure is preserved in the presence of a dielectric 
environment of the TMDC monolayer given in experiment, where it is placed on a sapphire substrate. We simulate the resulting screening effects with a macroscopic dielectric function as discussed in the Methods section. We assume that the TMDC layer is van der Waals-bound to the substrate with dielectric constant $\varepsilon=10$ and a distance of 3 \AA\ between the outer atomic layers of TMDC and substrate, which has been shown to be reasonable \cite{rooney_observing_2017,florian_dielectric_2017}. The resulting binding energies are $\big\{19.8,12.3,11.6,11.6,3.4,3.1\big\}$ meV.
Considering the optical selection rules discussed above, our microscopic theory thus predicts biexciton binding energies around 20, 12 and 3 meV to be observed in experiment.
These values are in very good agreement with the binding energies we obtained from fitting the differential absorption spectra as shown in Fig.~\ref{fig:exp_theo}.
The binding energies are only slightly overestimated by theory, most probably because screening by residual carriers is not taken into account.


Exchange interaction between electrons caused by their indistinguishability is at the heart of quantum mechanics and chemical bonding. Electrons and holes in a semiconductor on the other hand are approximately distinguishable, which generally results in weak electron-hole exchange effects. For example, in semiconductor quantum dots, the exciton fine structure due to electron-hole exchange is on the order of $\mu$eV, while in quantum wells no such effects have been observed. Our results show that atomically thin TMDC semiconductors are well-suited to study the fine structure of correlated Coulomb complexes experimentally as these materials combine large binding energies and relatively strong electron-hole exchange interaction. Biexcitons, as complexes that are fundamental to the nonlinear optical response of semiconductors, reveal a distinct fine structure that can be traced back to electron-hole exchange, along with local-field corrections to biexciton binding energies.
The fitting of experimental differential absorption spectra assuming three biexciton resonances, where the binding energies are very consistent at eight different pump wavelengths, strongly support the predictions of our microscopic theory quantitatively. Even more so as we can at the same time reproduce the complex differential line shape of the exciton resonance caused by biexciton correlations, which has not been reported before for TMDCs. 
This is consistent with the biexciton resonances to be spectrally located between exciton and trion as observed in monolayer MoSe$_2$ by two-dimensional coherent spectroscopy \cite{hao_neutral_2017}.
Previously, larger biexciton binding energies have been inferred from experimental observations of other stable many-body states in TMDCs \cite{mai_many-body_2014, you_observation_2015, plechinger_identification_2015, shang_observation_2015, sie_intervalley_2015, lee_identifying_2016, okada_observation_2017}.

\par 
We further demonstrate that profound insight into the relevant interaction processes can be gained by condensing the full microscopic theory into a simple configuration model for biexcitons.
From the observation that the simple model captures the structure of the full spectrum we conclude that 
the latter is 
essentially determined 
by the spin-selection rules of electron-hole exchange, while precise numbers can be obtained from microscopic calculations.
\par
The understanding of the biexciton fine structure opens a new field of research as the conclusions we draw are not limited to WSe$_2$ on sapphire substrate but are general for most vertical heterostructure based on semiconducting TMDCs. Even richer fine structures are expected for molybdenum-based materials due to the practically degenerate conduction bands, where more biexciton configurations can take part in the interaction. The configuration picture we present here can moreover be extended to consider electron-hole exchange in higher Coulomb complexes in TMDCs.
The biexciton fine structure does not rely on the existence of several inequivalent valleys as a single spin-degenerate conduction- and valence-band valley should have the same effect. Such a situation is given for thallium halides, where a configuration picture has already been applied to determine the effect of electron-hole exchange on the lowest biexciton state \cite{chung_theory_1983}. However, a biexciton fine structure has not been predicted there.
Neither has it been attempted to calculate material-realistic spectra for classical materials with strong biexciton signatures like CuCl. We conclude that reconsidering such materials by means of both modern computational and experimental methods could reveal previously undiscovered biexciton properties.

\par
In the light of our results, the question arises whether electron-hole exchange effects are carried over to TMDC nanostructures \cite{kern_nanoscale_2016, branny_deterministic_2017, palacios-berraquero_large-scale_2017}
or localized biexciton complexes \cite{he_cascaded_2016} both envisaged for quantum-optical applications. The biexciton fine structure in TMDC layers, as we describe it here, crucially depends on the existence of several four-particle configurations,
due to the continuous density of states provided by the band-structure valleys,
interacting via electron-hole exchange. 
If only a discrete density of states is realized in a
nanostructure, we expect just a single biexciton line as in semiconductor quantum dots. However, the microscopic understanding of TMDC nanostructures and their electronic states is still at its very beginning and remains a fascinating topic for future investigations.



\section{Methods}
\subsection{DCT Theory for biexcitons including electron-hole exchange}

The equations that describe coherent biexcitons in a pump-probe setup on a micropscopic basis are derived using the dynamics-controlled truncation (DCT) scheme \cite{axt_dynamics-controlled_1994, lindberg_chi_1994} that has been sucessfully used in the past to describe biexcitonic effects in quantum-well structures \cite{sieh_coulomb_1999}. This method allows to systematically account for nonlinear contributions to exciton dynamics up to a certain order in the exciting optical fields within the coherent limit. Biexcitonic contributions appear in third order ($\chi^{(3)}$ non-linearity) and the biexciton spectrum is governed by the homogeneous equations of motion for the four-particle correlations
\begin{equation}
    \begin{split} 
        B^{heh'e'}_{\bk,\bkp}(\bq):&=\Big<a^{h'}_{-\bkp-\bq}a^{e'}_{\bkp}a^{h}_{-\bk}a^{e}_{\bk+\bq} \Big>\\
        &-\Big<a^{h'}_{-\bkp-\bq}a^{e'}_{\bkp} \Big> \Big<a^{h}_{-\bk}a^{e}_{\bk+\bq} \Big>
        \label{eq:correlation}
    \end{split} 
\end{equation}
with electron and hole annihilation operators $a^{e}_{\bk}$ and $a^{h}_{\bk}$:
\begin{widetext}
\begin{equation}
  \begin{split}
\frac{d}{dt}B^{heh'e'}_{\bk,\bkp}(\bq)&=(\varepsilon^{h'}_{-\bk'-\bq}+\varepsilon^{e'}_{\bk'}+\varepsilon^{h}_{-\bk}+\varepsilon^{e}_{\bk+\bq})B^{heh'e'}_{\bk,\bkp}(\bq)\\
&-\frac{1}{\mathcal{A}}\sum_{\bqp\hbp\ebp}\Big(V^{e'\hbp h' \ebp}_{\bkp,\bkp+\bq-\bqp,\bkp+\bq,\bkp-\bqp}-U^{e'\hbp\ebp h'}_{\bkp,\bkp+\bq-\bqp,\bkp-\bqp,\bkp+\bq}\Big)B^{he\hbp\ebp}_{\bk,\bkp-\bqp}(\bq)\\
&-\frac{1}{\mathcal{A}}\sum_{\bqp\hb\eb}\Big(V^{e\hb h \eb}_{\bk+\bq,\bk-\bqp,\bk,\bk+\bq-\bqp}-U^{e\hb \eb h}_{\bk+\bq,\bk-\bqp,\bk+\bq-\bqp,\bk}\Big)B^{\hb\eb h'e'}_{\bk-\bqp,\bkp}(\bq)\\
&+\frac{1}{\mathcal{A}}\sum_{\bqp\hb\hbp}    V^{h'\hb h \hbp}_{\bkp+\bq+\bqp,\bk-\bqp,\bk,\bkp+\bq}B^{\hb e\hbp e'}_{\bk-\bqp,\bkp}(\bq+\bqp)\\
&-\frac{1}{\mathcal{A}}\sum_{\bqp\hb\ebp}\Big(V^{e'\hb h \ebp}_{\bkp,\bk-\bqp,\bk,\bkp-\bqp}-U^{e'\hb \ebp h}_{\bkp,\bk-\bqp,\bkp-\bqp,\bk}\Big)B^{\hb eh'\ebp}_{\bk-\bqp,\bkp-\bqp}(\bq+\bqp)\\
&+\frac{1}{\mathcal{A}}\sum_{\bqp\eb\ebp}    V^{e\ebp e' \eb}_{\bk+\bq,\bkp,\bkp+\bqp,\bk+\bq-\bqp}B^{h\eb h'\ebp}_{\bk,\bkp+\bqp}(\bq-\bqp)\\
&-\frac{1}{\mathcal{A}}\sum_{\bqp\hbp\eb}\Big(V^{e\hbp h' \eb}_{\bk+\bq,\bkp+\bq-\bqp,\bkp+\bq,\bk+\bq-\bqp}-U^{e\hbp \eb h'}_{\bk+\bq,\bkp+\bq-\bqp,\bk+\bq-\bqp,\bkp+\bq}\Big)B^{h\eb \hbp e'}_{\bk,\bkp}(\bq-\bqp)\,.
    \label{eq:eom}
    \end{split} 
\end{equation}
\end{widetext}
$\mathcal{A}$ denotes the crystal area. Eq.~(\ref{eq:eom}) represents a generalized four-particle Schr\"odinger equation for zero total momentum in reciprocal space including arbitrary band structures $\varepsilon^{\lambda}_{\bk}$ and Coulomb matrix elements $V^{\lambda_1\lambda_2\lambda_3\lambda_4}_{\bk+\bq,\bkp-\bq,\bkp,\bk}$ in the c-v picture as usually obtained from first-principle calculations. 
The six ``direct'' interaction terms with matrix elements $V$ and momentum transfer $\bqp$ are analogous to Coulomb interaction between the four particles in a hydrogen molecule. 
The additional terms describe electron-hole exchange, which is beyond the picture of biexcitons as semiconductor counterpart to hydrogen molecules.
As the Coulomb interaction is spin-diagonal, only the electron-hole exchange terms couple biexcitons with different spin combinations. 
We consider the two highest conduction and two lowest valence bands of a TMDC semiconductor here, hence $e$ and $h$ label the respective spin index. Otherwise there would be additional exchange terms between different conduction and 
different valence bands in Eq.~(\ref{eq:eom}). Note that the electron-hole exchange interaction is described by unscreened Coulomb matrix elements $U$, see \cite{sham_many-particle_1966, denisov_longitudinal_1973,qiu_nonanalyticity_2015}, while $V$ denotes matrix elements screened by carriers in occupied bands and the dielectric environment of the TMDC layer.
The first three lines of Eq.~(\ref{eq:eom}) describe two independent excitons containing the exciton-exciton continuum edge as lowest eigenstate.

\par

As those TMDC biexcitons visible in pump-probe experiments are hosted in the K and K' valleys of the Brillouin zone, we limit our single-particle basis to these valleys, where we carefully converge our results with respect to the radius in $\bk$-space around K and K' and to the resolution of the used Monkhorst-Pack mesh. More details on the convergence are given in the Supporting Information. To obtain biexciton spectra numerically, we have to diagonalize the homogeneous equation of motion (\ref{eq:eom}) in matrix form set up in a basis defined by the quantum numbers $\left\{\bk,\bkp,\bq,e,h,e',h'\right\}$. Since the resulting matrix can amount up to several Terabytes depending on the used $\bk$-mesh, we utilize an iterative Krylov space method as contained in the SLEPc package \cite{hernandez_slepc:_2005} for the PETSc toolkit \cite{balay_petsc_2016} to obtain only the lowest eigenvalues.
Due to the numerical discretization of momentum space, there is an uncertainty of binding energies of $\pm 1$ meV.

\subsection{Material-realistic description of Coulomb interaction}

We combine the above many-body theory of biexcitons with ab-initio methods providing material-realistic band structures and bare as well as screened Coulomb matrix elements on a G$_0$W$_0$ level as input for the equations of motion (\ref{eq:eom}). More details on the G$_0$W$_0$ calculations are given in Ref.~\citenum{steinhoff_influence_2014}. The valence- and conduction-band splitting caused by spin-orbit interaction is described along the lines of \cite{liu_three-band_2013,steinhoff_influence_2014} including first- and second-order effects. To take into account dielectric screening by the environment, we use the \textit{Wannier function continuum electrostatics} (WFCE) approach described in \cite{rosner_wannier_2015,florian_dielectric_2017} that combines a continuum electrostatic model for the screening with a localized description of Coulomb interaction. A parametrization of Coulomb interaction in monolayer WSe$_2$ in terms of a localized basis $\ket{\alpha}$ consisting of Wannier functions with dominant 
W-d-orbital character is provided in Ref.~\citenum{steinhoff_exciton_2017}. We obtain Coulomb matrix elements in Bloch-state representation by a unitary transformation
\begin{equation}
  \begin{split}
& U^{\lambda_1\lambda_2\lambda_3\lambda_4}_{\bk+\bq,\bkp-\bq,\bkp,\bk} = \\
 &\sum_{\alpha\beta\gamma\delta}
  \big(c_{\alpha,\bk+\bq}^{\lambda_1}\big)^*\big(c_{\beta,\bkp-\bq}^{\lambda_2}\big)^* \, c_{\gamma,\bkp}^{\lambda_3} c_{\delta,\bk}^{\lambda_4} \,\,
 U_{\bq}^{\alpha\beta\gamma\delta}
    \label{eq:unitary_transform}
    \end{split} 
\end{equation}
with coefficients $c_{\alpha,\bk}^{\lambda}$ that connect the localized and the Bloch basis on a G$_0$W$_0$-level as described in \cite{steinhoff_influence_2014}. While for the direct Coulomb interaction density-density-like contributions $V_{\bq}^{\alpha\beta\beta\alpha}$ are dominant, in turns out that a proper description of electron-hole exchange additionally requires matrix elements that are exchange-like in the local representation:
\begin{equation}
U_{\bq}^{\alpha\beta\gamma\delta}=U_{\bq}^{\alpha\beta\beta\alpha}\delta_{\alpha\delta}\delta_{\beta\gamma}+U_{\bq}^{\alpha\beta\alpha\beta}\delta_{\alpha\gamma}\delta_{\beta\delta} \,.
\label{eq:exchange}
\end{equation}
The density-density-like matrix elements are momentum-dependent and thus nonlocal in real space with a characteristical $1/q$-behavior for small momenta. A numerical analysis shows that they yield the main contribution to the longitudinal-transverse exciton splitting shown in Fig.~2 in the Supporting Information that leads to a light-like dispersion of the upper exciton branch. The exchange-like matrix elements on the other hand are practically momentum-independent and correspond to local interaction processes. They alone are responsible for the almost momentum-independent exciton blue shift, which is therefore termed a local-field effect. Unlike the density-density-like matrix elements, the exchange-like matrix elements are not obtained via WFCE but fit directly with a constant value.
\par

The exchange effects can also be understood by analyzing Coulomb matrix elements like $U^{e\hb \eb h}_{\bk+\bq,\bk-\bqp,\bk+\bq-\bqp,\bk}$ in Eq.~(\ref{eq:eom}) for $\bk$ and $\bk-\bqp$ in the vicinity of K and K' and small exciton momenta $\bq$. To this end, we resort to the $d_{m=0}$- and $d_{m=\pm2}$-
orbital character of Bloch states in the K- and K'-valleys as obtained 
analytically from a simple lattice model \cite{xiao_valley-contrasting_2007, xu_spin_2014}. It turns out that index combinations $U^{\alpha\beta\beta\alpha}$ lead to intra- and inter-valley exchange terms that scale linearly with $|\bq|$, while combinations $U^{\alpha\beta\alpha\beta}$ yield intra-valley terms that have constant and quadratic contributions.
 A similar discussion of electron-hole exchange in much detail has been led by Qiu et al. for monolayer MoS$_2$ \cite{qiu_nonanalyticity_2015}, with the difference that they use a plane-wave description of Coulomb interaction.
In monolayer WSe$_2$ the exchange-like matrix elements, if normalized to a unit-cell area, amount to 0.39 eV for interaction between $d_{m=0}$- and $d_{m=\pm2}$-orbitals and to 0.19 eV for interaction among $d_{m=\pm2}$-orbitals. We find that matrix elements involving three different orbitals that would in principle additionally contribute to the above-mentioned effects are negligible with respect to the other ones.

\subsection{Simplified model of electron-hole exchange}

Eq.~(\ref{eq:eom}) constitutes a four-particle Hamiltonian whose eigenstates describe biexcitons with vanishing total momentum. The Hamiltonian can be split into a part without electron-hole exchange $H_0$, a part including nonlocal exchange $H_\textrm{U}$ and a local-field part $H_{\textrm{LF}}$ according to the Coulomb matrix elements $V$, $U^{\alpha\beta\beta\alpha}$ and $U^{\alpha\beta\alpha\beta}$ involved, see Eq.~(\ref{eq:exchange}). The four-particle configurations 
\\
$\Big\{\ket{\textrm{B}^{\uparrow\Uparrow\uparrow\Uparrow}_{\textrm{KK}}(0)},\ket{\textrm{B}^{\uparrow\Uparrow\downarrow\Downarrow}_{\textrm{KK'}}(0)},\ket{\textrm{B}^{\downarrow\Downarrow\uparrow\Uparrow}_{\textrm{K'K}}(0)}  ,$ $\ket{\textrm{B}^{\uparrow\Downarrow\downarrow\Uparrow}_{\textrm{K'K}}(-K)}, \ket{\textrm{B}^{\downarrow\Uparrow\uparrow\Downarrow}_{\textrm{KK'}}(K)},\ket{\textrm{B}^{\downarrow\Downarrow\downarrow\Downarrow}_{\textrm{K'K'}}(0)} \Big\}$ are eigenstates of $H_0$ with eigenenergies $E$.
These configurations with carrier spins as upper indices are superpositions of four-particle states $B^{heh'e'}_{\bk,\bkp}(\bq)$ given by the quasi-momenta $-\bk$, $\bk+\bq$, $\bk'$ and $-\bk'-\bq$ as evident from Eq.~(\ref{eq:eom}). The configurations can also be seen as superpositions of two-exciton states with different exciton momenta. This is expressed by the lower configuration indices denoting the valleys where the excitons reside, while the dominant exciton momentum $\bq$ is given in brackets. 
Electron-hole exchange introduces off-diagonal matrix elements $\bra{B_i}H_\textrm{U}+H_{\textrm{LF}}\ket{B_j}=U_{ij}+\Delta_{\textrm{LF},ij}$. Taking into account the electron and hole spins in the basis configurations, we can determine those exchange matrix elements that do not flip carrier spins and are thus non-zero. As a simplification, we assume that all non-zero exchange matrix elements have the same values $U$ and $\Delta_{\textrm{LF}}$, respectively. Then the full Hamiltonian in the 
subspace of bound biexcitons is given as a 6x6 matrix by
\arraycolsep=2.5pt\def\arraystretch{0.5}
\begin{equation}
    \begin{split} 
  H= (E+2U)\mathbb{1}+U&
\left(
\begin{array}{cccccc}
2 & 1 & 1 & 1 & 1 & 0 \\
1 & 0 & 0 & 0 & 0 & 1 \\
1 & 0 & 0 & 0 & 0 & 1 \\
1 & 0 & 0 & 0 & 0 & 1 \\
1 & 0 & 0 & 0 & 0 & 1 \\
0 & 1 & 1 & 1 & 1 & 2
\end{array}
\right) \\ +2\Delta_{\textrm{LF}}&
\left(
\begin{array}{cccccc}
1 & 0 & 0 & 0 & 0 & 0 \\
0 & 1 & 0 & 0 & 0 & 0 \\
0 & 0 & 1 & 0 & 0 & 0 \\
0 & 0 & 0 & 0 & 0 & 0 \\
0 & 0 & 0 & 0 & 0 & 0 \\
0 & 0 & 0 & 0 & 0 & 1
\end{array}
\right)  \\ &=H_1+H_{\textrm{LF}}\,.
        \label{eq:matrix}
    \end{split} 
\end{equation}
Biexciton energies $E$ without electron-hole exchange appear as diagonal entries together with twice the exchange matrix element $U$, as each biexciton state allows for at least two different exchange processes between like-spin electron-hole pairs. 
The second matrix takes care of all possible interaction processes between the biexciton configurations and the final one accounts for the local-field shifts of those states that contain two like-spin excitons, see Fig~\ref{fig:eh_exchange}~(a), which excludes configurations 4 and 5. To obtain a clear picture, we diagonalize the matrix $H_1$ without local-field contributions, which we include afterwards using perturbation theory. The eigenenergies of $H_1$ are given by $E_i=\left\{E,E+2U,E+2U,E+2U,E+4U,E+6U \right\}$ and the corresponding eigenstates $\ket{\Phi_i}$ are
\\
$1/\sqrt{6}(1,-1,-1,-1,-1,1)^T$, $1/2(0,1,1,-1,-1,0)^T$, $1/\sqrt{2}(0,1,-1,0,0,0)^T$,
\\
$1/\sqrt{2}(0,0,0,1,-1,0)^T$,$1/\sqrt{2}(1,0,0,0,0,-1)^T$ and $1/\sqrt{12}(2,1,1,1,1,2)^T$.
The first-order energy correction to the lowest eigenenergy is $\bra{\Phi_1}H_{\textrm{LF}}\ket{\Phi_1}=4/3 \Delta_{\textrm{LF}}$.

\subsection{Sample characterization}
Based on previous studies of similarly prepared samples \cite{singh_trion_2016}, we consider that the sample is $n$-doped. The electronic structure of the sample is characterized by exciting photoluminescence (PL) with a helium neon laser (632.8~nm), and measuring the differential absorption, using the output of a titanium-sapphire laser tuned across the A exciton and trion resonances of the sample. In both cases, we obtain spectra with clearly pronounced inhomogeneously broadened signatures of the neutral exciton and the two overlapping trions. More details are given in the Supporting Information.
\subsection{Pump-probe experiment}
The experiments are carried out by splitting the output of a titanium-sapphire laser 
to
a pump and a probe beam. The output pulses have a spectral full width at half maximum (FWHM) of 15~nm. Their spectra are shaped in two grating-based pulse-shapers using an amplitude mask to a spectral width of about 0.7~nm FWHM, corresponding to a duration of about 1~ps FWHM in time. The intensity of both beams is modulated by passing them through acousto-optical modulators driven at about 3~MHz with a difference of a few kHz.
Polarization control is carried out by linear polarizers and subsequent quarter-wave plates.
The beams are overlapped spatially in a beamsplitter cube and focused on the sample by a 100x microscope objective to a spot size of 2~$\mu$m in diameter. 
Intensity dependent studies are performed to ensure that the data taken at $\approx 10$ $\mu$W are in the $\chi^{(3)}$ regime and to avoid excessive heating. 
After passing through the sample, the pump part is filtered out by a monochromator. The differential transmission is detected by a silicon detector coupled to a lock-in amplifier set to the difference frequency of the pump and probe modulations. Assuming no purely reflection-related resonances, this can be directly correlated to the negative differential absorption.





\section{Acknowledgement}
This work was supported by Deutsche Forschungsgemeinschaft (DFG) within CRC 1558 and RTG 2247.  A. W. A. acknowledges funding from DFG via grant No. AC290-2/1. The spectroscopic experiments were jointly supported by NSF DMR1306878 (A.S.) and NSF MRSEC program DMR-1720595 (K.T.). X.L. gratefully acknowledges support from the Welch foundation (F-1662) and the Alexander von Humboldt foundation, which facilitated the collaboration with TU-Berlin. A.St. and M.F. would like to acknowledge Paul Gartner for fruitful discussions. We thank Gunnar Sch\"onhoff, Malte R\"osner and Tim Wehling for providing material-realistic band structures and bare as well as screened Coulomb matrix elements. 
\section{Competing interests}
The authors declare that they have no competing financial interests.
\section{Author contributions} 
A.~St. and M.~F. performed analytical and numerical calculations of the biexciton spectra. A.~S., K.~T., and M.~K. designed and performed the experiments. K.~T. exfoliated the sample. N. O., A. S., A. W. A. and S. H. analyzed the data. N. O., A. St., M. F., and A. S. prepared the manuscript. U.~W., F.~J., and X.~L. initiated and coordinated the project. All authors contributed to the discussion and the writing of the manuscript.
%

\end{document}